# Impact Of Urban Technology Deployments On Local Commercial Activity

Stanislav Sobolevsky[1,4], Ekaterina Levitskaya[1], Henry Chan[2], Shefali Enaker[2], Joe Bailey[2], Marc Postle[2], Yuriy Loukachev[3], Melinda Rolfs[3], Constantine Kontokosta[1,5]

[1]Center For Urban Science And Progress, New York University, Brooklyn, NY, USA
[2] Future Cities Catapult, London, UK
[3] Center for Inclusive Growth - Data Science, Mastercard, Purchase, NY, USA
[4]Institute Of Design And Urban Studies of The National Research University Of Information Technologies, Mechanics And Optics, Saint-Petersburg, Russia
[5]Department of Civil and Urban Engineering, New York University, Brooklyn, NY, USA

While smart city innovations seem to be a common and necessary response to increasing challenges of urbanization, foreseeing their impact on complex urban system is critical for informed decision making. Moreover, often the effect of urban interventions goes beyond the original expectations, including multiple indirect impacts. The present study considers the impact of two urban deployments, Citi Bike (bike sharing system) and LinkNYC kiosks, on the local commercial activity in the affected neighborhoods of New York City. The study uses anonymized and aggregated insights provided through a grant from the Mastercard Center for Inclusive Growth in order to provide initial data-driven evidence towards the hypothesis that proximity of Citi Bike stations incentivizes local sales at eating places, while LinkNYC kiosks help people, especially visitors, to navigate local businesses and thus incentivize commercial activity in different business categories.

**Introduction**

Modern cities face non-trivial sustainability and operational challenges due to growing populations, ageing infrastructure, and the increasing complexity of urban systems. While smart city innovations have become an attractive tool to try to address these challenges, urban decision-makers must be able to justify such innovations by understanding and evaluating their impact, a process which is complicated by the nature and diffusion of technology in the urban environment.

The growing availability of big urban data, capturing various aspects of urban activity, provides an unprecedented opportunity to model urban complexity and assess the impact of discrete technological interventions. Over the past few years, datasets on urban activity such as cell phone connections [1-8], GPS readings [13], social media [16-19], as well as various sensor data [14, 15], have been explored as useful proxies for urban mobility and other types of urban behavior.

One of the most relevant data sources for understanding commercial activity is the data on consumer spending behavior. This type of data provides a glimpse into the dynamics of economic and purchasing trends at high spatial and temporal resolution and have been successfully applied to sensing urban dynamics and socio-economic performance [9-12] across neighborhoods, cities, and regions.

Aggregated and sometimes normalized (into relative scores) datasets reflecting various types of the cumulative spending activity of urban neighborhoods can serve as a useful proxy for many sorts of urban economic studies. In this paper, we leverage insights provided by Mastercard Retail Location Insights [20] to perform initial evaluation of the economic impacts of urban innovations on local commercial activity. We focus on the recent deployments of the Citi Bike bike-sharing system and LinkNYC, digital kiosks providing free gigabit WiFi, across New York City (NYC).

Bike-sharing systems have gained a lot of popularity in recent years, as biking has been associated with a host of public benefits as a healthy and environmentally friendly transportation alternative [21-25]. Rapidly growing bike sharing systems all over the world have attracted a lot of attention by urban analysts and researchers, including the study of their economic, social, and environmental impacts. In New York City's congested Midtown Manhattan, typical trips between 1-1.5 miles are more than 5 minutes faster and $10 cheaper by Citi Bike than by taxi [26]. In Montreal, bicycle sharing systems have been shown to increase property values for multifamily housing by approximately 2.7% within an 800 meter buffer of individual docking stations [27]. Investments in bicycle infrastructure create nearly twice as many direct, indirect and induced jobs per dollar as typical road projects [28]. Existing literature also highlights time savings, improvements in public health, social inclusion and increase in public safety. Fully 70% of Capital Bikeshare riders (Washington D.C.) choose bikeshare as they believe that it is the quickest and easiest way to get to their destination [29]. Bicycling to work has been found to decrease risk of mortality [30]. The authors of a Deloitte Smart Mobility report suggest that a national total for 171 statistical city metropolitan areas in the United States amounts to $2.6 billion annually in indirect savings based on lower road construction costs, reduced accidents, and lower carbon dioxide emissions [31].

Some studies have also investigated the links between bicycle sharing and local commercial activity. In Minneapolis, a study of Nice Ride system find correlations between how bike share activity increases with the number of eating places within a ⅛ mile walk of the bike share station; the researchers define "station activity" through three measures: the sum of trip origins and trip destinations; the number of trip origins, and the number of trip destinations [32].

The benefit of providing free public Wi-Fi has also become increasingly acknowledged by city administrations. Another recent deployment in NYC, LinkNYC kiosks, in addition to a free access to Internet, also includes access to map navigation, phone calls and online access to public services. The sidewalk interactivity provided by LinkNYC kiosks impacts the overall attractiveness of public spaces and businesses within them.

A recent deployment of LinkNYC kiosks has been mentioned in research studies related to the examination of smart city initiatives. Several studies discuss the issue of privacy often associated with "open data platforms" that smart city administrations use to increase civic participation and expand public services to residents [33, 34]. Another study cites LinkNYC as an example of a "hyperlocal smart city experience" and contrasts it with other smart city innovations which usually target a metropolis scale rather than an individual scale. The authors of this study argue that more case studies of smart city innovations which articulate a "greater level of sociability and care for the individual" are lacking [35].

While LinkNYC kiosks include pervasive displays and there is a lot of research on the real-time location-based marketing techniques, including the International Symposium on Pervasive Displays [36-37], this accounts only for a commercial aspect of the project, while the uniqueness of LinkNYC kiosks stems from their proposed public benefit, including access to a 911 button, public services, free Wi-Fi, free navigation services, and free calls. Although the LinkNYC project has been announced as one of the ways to help to bridge the digital divide within the city [38], several studies on social justice question it. One study deems LinkNYC a "public Wi-Fi" which has been developed as a "profit-driven" infrastructure focused on generating advertising revenue [39]. The author of the study contrasts it with the Red Hook Wi-Fi project which has been developed as a community-based initiative in Brooklyn and doesn't have a corporate, for-profit component to it, as does LinkNYC which is managed by a consortium of corporations. A similar discussion is found in another study on digital divide and inequality where LinkNYC is mentioned as one of the smart city projects aimed at bridging the digital divide [40]. Finally, another study related to social justice suggests that smart city projects like LinkNYC promote "'smart enclaves' for exceptional citizens" and do not encourage true inclusiveness [41].

We use scored aggregated commercial activity insights provided through a grant from the Mastercard Center for Inclusive Growth [20] to estimate economic activity around Citi Bike stations in NYC and neighborhoods in order to evaluate the hypothesis that Citi Bike station deployments might incentivize purchases at eating places in the area. It can be argued that by encouraging more purchases at eating places in the area (food trucks, delis, small shops with food and beverages or a walk-in cafes, etc.), Citi Bike stations help to promote small businesses and thus contribute to an inclusive economic growth and a diverse commercial profile of the area. Similarly, we consider the hypothesis that deployment of LinkNYC kiosks that are supposed to help pedestrians to navigate the city increase the exposure of local businesses to the passersby, and will therefore result in incentivizing commercial activity in various business categories across a given place. Studying the impact of these two urban innovation systems - Citi Bike and LinkNYC - on local commercial activity will provide a data-driven analysis to support the economic rationale for the deployment of these urban innovations.

**Citi Bike deployment**

Citi Bike, launched in May 2013 in New York City, is the largest bike share system in the United States [42]. The system was backed by Citibank as the title sponsor and Mastercard as the Preferred Payment Partner. Citi Bike is available for use 24 hours per day, 7 days a week and during all seasons. Riders use credit cards to rent a bike through three different plans (day pass, 3-day pass, annual membership). Additional web applications provide the opportunity to track the number of available bikes at a given station in the real time. New York City Department of Transportation described choosing sites for original 600 Citi Bike stations as "one of the most participatory planning processes ever undertaken in New York" [43]. This effort included stakeholder meetings, an innovative online portal for gathering input (which was then replicated for planning purposes in other U.S. cities), demonstration events and community planning workshops. Currently, there are 609 stations and around 10,000 bikes in Manhattan, Brooklyn, Queens and Jersey City [44].

One of the most obvious impacts of Citi Bike stations is on urban mobility and associated health benefits due to additional physical exercise. However, beyond those direct impacts, the potential indirect impacts could be anticipated through an additional level of human activity around the stations promoted by Citi Bike. The impacts on the immediate surroundings might involve local businesses. Citi Bike riders are potential customers for local businesses around the stations, especially for eating places due to the physical exercise involved in cycling. Several studies confirm the link between cycling as a travel mode and consumer spending behavior, with cyclists spending similar amounts or more than those traveling in automobiles [45]. Bikeshare users in Washington D.C. reported making new or induced trips to commercial districts because of bikeshare stations available there, and as a results spending more money because of it [46]. Another study suggests that bike lanes act as a catalyst for local economic activity [47].

**LinkNYC deployment**

LinkNYC is a communications network that was launched in New York City in 2015 and replaced public payphone structures with free interactive kiosks [48]. The infrastructure of LinkNYC is owned and maintained by CityBridge, a consortium of experts in technology, advertising, connectivity and user experience from Qualcomm, Intersection, CIVIQ Smartscapes companies and the City of New York. LinkNYC is intended to generate its own revenue through advertising, sponsorships and partnerships, as well as more than half of a billion dollars for the City of New York [49]. LinkNYC kiosks are free to use and do not only provide broadband access (many cities have seen Wi-Fi deployment in public spaces), but they also offer additional options for interactivity through transit or emergency announcements, interactive maps of the area, local ads and directions to local businesses and services. On the economic side, LinkNYC is intended to help raise public awareness of the local offerings, increasing the accessibility of and sales at local businesses. Since 2015, in total 7,500 kiosks have been deployed in New York City [49].

Similarly to Citi Bike stations, LinkNYC kiosks impact their immediate surroundings by providing an additional sidewalk interactivity option which promotes greater human activity around the kiosks, from which local businesses may also benefit. Due to a local navigation option and advertising displays, LinkNYC kiosks enhance the awareness of the local area and may convert pedestrians into potential customers for local businesses. Existing literature, similarly to the effect of cyclists on local economic performance, point to the evidence that pedestrian friendly environments increase retail sales performance and local business activity [50].

**Commercial Activity Insights**

Commercial activity insights considered for the project are available through the Mastercard Retail Location Insights tool [20], which provides aggregated commercial activity performance scoring for the areas of interest. The MRLI Scores on a monthly basis show for a selected geographic area (census block, tract, zip code, country or state), business category of interest (total retail , apparel, dining, total retail excluding dining and small businesses), and characteristic of interest (rankings for total sales, transaction count, average receipt amount, growth rate and stability), which quantile of the country-wide distribution does the given characteristic of the selected area belongs to. Although the ranking score reported by MRLI is a relative quantity, by knowing the overall log-normal shape and parameters of the country-wide distribution of commercial activity (sales volumes) reported by Census [51] we can convert it to a linear proxy of the commercial activity by applying an inverse cumulative distribution transform. Resulting values are still an approximation and do not represent the actual scale of the payment card activity in the local businesses. Those values are used exclusively on a relative scale for the purpose of comparative analysis over different areas and timeframes. Also, scores are not provided by MRLI for the areas not meeting compliance guidelines, limiting the spatial coverage and the granularity of our analysis. However, MRLI insights still serve as a unique proxy for commercial activity around the city.

For locations of Citi Bike stations and LinkNYC kiosks, we use data from the Citi Bike System Data [52] and NYC Open Data portal [53]. Citi Bike publishes large open datasets on its website, including daily ridership, membership data and trip histories [52]. Trip histories data includes trip duration (seconds), start and stop time and date, start and end station names, station ID, station lat/long, bike ID, user type (customer = 24-hour pass or 7-day pass user; subscriber = Annual Member), gender and year of birth. The New York City Open Data portal provides data on LinkNYC kiosk locations, including Link ID, borough, community board, council district, latitude, longitude and project status [53].

**Methodology**

The basic approach for our commercial impact assessment is comparing the level of commercial activity within the area of interest over comparable periods (i.e. same months of different years) before and after the deployment. However, even if commercial activity does change after the deployment, it is not by itself sufficient to claim that the deployment was indeed the cause of such a change. The latter can still be a side effect of other local factors of influence, including the overall trends in the entire study area. The common way of addressing this issue is through counterfactual analysis, comparing impacts of the intervention against a baseline (a "without intervention" scenario) [54]. Counterfactual analysis is frequently used in social and economic science research studies [55-57]. Evaluating change over time in the variable of interest for the deployment area (treatment group), compared to the change over time for the control area is often referred to as difference-in-differences approach [58].

General factors causing the city-wide or neighborhood-wide trends are relatively straightforward to control for by comparing the deployment areas against the nearby control areas, those identified as similar to the deployment area, but for the deployment of interest. This type of spatial counterfactual analysis is our primary method of impact evaluation. The statistical significance of the difference between the distributions of the parameters of the commercial activity dynamics within the deployment area and within the control area are evaluated using the above difference-in-differences approach as well as the Kolmogorov-Smirnov (KS)-test [59].

However, no single control area in a complex urban system would have exactly the same characteristics as the deployment area does. This leaves a possibility for the observed difference between the dynamics of commercial activity within the deployment area in comparison with the dynamics in the control area (even if statistically significant) to be caused by other local factors not necessarily related to the deployment. Such local factors could be apparently affecting the deployment area, but might not necessarily be present in the control area – in a complex urban system it is hardly feasible to control for all such factors.

Because of that, an additional temporal test is useful to further evaluate a causal relationship. As it is less likely (although still possible) that additional local factors influence the urban system at exactly the same time as the intervention of interest, it could be helpful to compare the dynamics of the local characteristics in the deployment area before and after the deployment occurred. If the same effect was observed preceding the deployment, then it is likely that the deployment is not the cause (e.g. if pedestrian counts grow in the affected area after the deployment, but also do so over some time before, it is evident that there is some other local cause for such growth). For example, one could compare the change in quantities of interest over a year after the deployment against a year before, on one hand, with a change of the quantities one year before the deployment against the preceding year. We further refer to this temporal counterfactual analysis as "placebo test" [60-63], while it also sometimes referred to as a difference-in-difference-in-difference (DDD) approach [64, 65].

An additional possibility is to utilize a distance decay approach in order to verify whether commercial activity in the identified areas gradually decreases while moving further away from the deployment locations, as one would expect in case the deployment would be the cause. Distance decay assessment is widely used to assess whether the phenomenon in question decreases with a spatial distance from the source [66-69]. Applying this approach for the purpose of the present research is subject to the availability of the data of sufficient spatial granularity and we see it as a possibility for the future studies.

**Impact of Citi Bike stations on the nearby commercial activity**

Another hypothesis we want to validate, as seen from research studies on bike sharing in other localities [27], is that Citi Bike deployments in New York City incentivize nearby commercial activity, in particular activity for eating places. In order to understand the impact of Citi Bike station deployments on the nearby commercial activity, we use the anonymized and aggregated insights of scored commercial performance of different areas (counties, zip codes, census tracts, census block groups and census blocks).

First, we look at the changes in the activity of local businesses around the deployment neighborhoods at different spatial scales over a year after the deployment, compared to the performance of the same area during the previous year (in order to exclude any seasonality effect). We also perform a comparative analysis with the dynamics of the nearby control areas where deployment did not occur (in order to control for the overall economic trends in the area).

Studying the impact of Citi Bike deployments in NYC, we focus on the borough of Brooklyn, as many of the deployment locations in Manhattan are among the top percentile for eating, which makes it more difficult to track fluctuations. For each phase of the deployment happening in July 2013, September 2015, and September 2016, we aggregate all zip code areas containing the deployed stations into the "deployment area" and include nearby zip codes not containing the deployed stations at that time into the "control area" (see the visualization of the deployment areas all over the city on Figure A1 of the Appendix and deployment zones in Brooklyn with the adjacent zip codes on Figure A3). For each of the selected locations, we assess the pre/post change in the overall transactional activity, comparing the year prior to the year following the year after the deployment.

**Table 1**. Estimated overall change for the transactional activity (transaction count) for eating places in Brooklyn after three phases of Citi Bike deployment.

| Deployment phase | Estimated relative change in the deployment area, % | Estimated relative change in the control area, % |
|---|---|---|
| 2013 | 0.546 ± 0.144 | -0.107 ± 1.034 |
| 2015 | 0.446 ± 0.615 | 0.004 ± 0.480 |
| 2016 | 0.209 ± 0.011 | -0.017 ± 0.489 |

From Table 1 we see that for all three stages of the deployment, the estimated transactional activity of the affected eating places grows around 0.2-0.5% over the post-deployment period, while control areas demonstrate mixed performance from a slight decrease to no change. However, standard deviations of the growth rates are relatively high compared to the averages, so statistical significance of the pattern requires further evaluation. Comparison of the distributions of growth rates within the deployment area and the control sample using Kolmogorov-Smirnov test gives a p-value of 7.24%. This way, we can confirm the statistical significance of the pattern only at a 90% confidence level rather than the standard 95% confidence, still leaving a small chance that the observed pattern is simply a random coincidence - one should not reject the null-hypothesis that the growth rates in the sample zip codes are randomly obtained from the same distribution, as the growth rates observed within the control areas. Also, the pattern of the deployment areas growing faster than control areas is not confirmed when analyzing the dynamics of the overall sales volumes, rather than transaction counts. One challenge in identifying a signal is that zip code areas are quite large and not all the businesses located within the same zip code are actually affected by a given deployment, thus producing spatially heterogeneous impacts across the zip code. Further we consider the impact at a more fine-grained spatial scale.

Census block areas are small enough (150x150m on average across NYC depending on the population density) to account for the proximity of the nearest station being deployed: while some blocks within the deployment area might have the stations deployed within them or nearby, others might have no stations within a reasonable proximity and can serve as a control area, even though they are generally located within the same neighborhood (which is even more beneficial, as such areas might be closer in context to the deployment blocks) having multiple deployments happening around it. This fine-grained analysis is not possible, however, as the MRLI tool restricts users from accessing information on individual business performance, with only the blocks having more than five businesses of the considered category being scored.

Below we consider the case of another city closely attached to New York: Jersey City in the State of New Jersey, which had multiple Citi Bike deployments in September 2015. The city is selected as a smaller deployment area where locating all the affected census blocks is easier. Several census blocks directly adjacent to the two major deployment locations – Grove St. and Journal Square Path train stations – are available (Figure A4). All of these were included into the corresponding deployment areas. We also found several census blocks in downtown Jersey City and nearby Hoboken with available data and no Citi Bike (nor Hudson Bike Share in case of Hoboken) stations deployed within close proximity. The screenshots from the MRLI tool visualizing the levels of commercial activity before and after the Grove St. Citi Bike stations were deployed are found on the Figure 1.

From Table 2 we can see that, again, deployment areas show considerable estimated growth of the eating places commercial activity over a year after the deployment. But now the average increase in deployment areas is already around 3.6% according to the Table 3, which is considerably higher than the modest increase we saw at the zip code scale. This is perhaps due to the fact that now the considered deployment areas are much smaller and all their businesses are in direct proximity from the bike stations, which was clearly not the case in larger zip code areas.

In comparison, the control area (all available census blocks in Jersey City and nearby Hoboken not directly adjacent to Citi Bike stations, although worth mentioning that there is only a handful of such) does not show any significant growth (and even a slight decrease), as reported in Table 3.

**Table 2.** Census Block GeoID's with available data on eating places adjacent to Grove St. and Journal Sq. Citi Bike deployments and the commercial activity (total sales) growth rate in them over 12 months after the deployment (September 2015 - August 2016) versus similar period one year before (September 2015 - August 2016) and also the growth rate one year before the deployment (September 2015 - August 2016 versus September 2014 - August 2015) for the placebo test.

| Census block GeoID | Estimated relative change after the deployment, % | Estimated relative change one year before, % |
| --- | --- | --- |
| 340170009022007 | 1.3295 | -0.1958 |
| 340170019001002 | 2.829 | -4.5818 |
| 340170019001003 | 1.3909 | -0.9560 |
| 340170070003007 | 1.7859 | 1.3274 |
| 340170070003010 | 10.7856 | 5.5787 |

In order to verify the statistical significance of the pattern, we compare the distribution of the commercial activity growth factors in both deployment areas with the distribution of the commercial activity growth factors in the control area. This time, a Kolmogorov-Smirnov test gives a p-value of 0.6861%, so the hypothesis that the distribution of the factors in the deployment area follows the same distribution as for the factors in the control area can only be rejected at a confidence level of 99%.

The difference-in-differences approach [65] provides another way to evaluate the significance of the above pattern through a linear regression

$$Y \sim b_0 + b_1 T + b_2 S + b_3 TS,$$

where Y is the quantity of interest within a given location (transaction count or sales volume) over a considered period of time (month), $b_{0,1,2,3}$ are the regression coefficients, T is a binary dummy variable being 1 is the period is after the deployment and 0 otherwise, S is a binary dummy variable being 1 is the location is within the deployment area and 0 otherwise, while TS is a another dummy variable being a product of T and S (1 is location is within the deployment area and the period is after the deployment and 0 otherwise). If $b_3$ happens to be substantially different from zero (e.g. judging by its confidence interval or the corresponding p-value), then the spatio-temporal scope of the deployment, i.e. having both conditions together - location is within the deployment area and the period is after the deployment - has a statistically significant impact on the quantity of interest. For the eating places commercial activity around the above deployments in Jersey City above, the p-value for $b_3$ is 2.74% for transactional activity and 0.03% for sales volume, i.e. below a standard 5% significance threshold, confirming the statistical significance of the relation between the commercial activity and the spatio-temporal scope of the deployment.

**Table 3.** Change in commercial activity (eating places) around recent deployments in Jersey City (Sept, 2015) versus the control sample, non-food-related sample and placebo test.

| Deployment phase | Estimated relative change in sales volume for the deployment area, % | Estimated relative change in transactional activity for the deployment area, % |
|---|---|---|
| Deployment | 3.624 ± 3.621 | 2.723 ± 1.175 |
| Control | -0.894 ± 0.681 | 0.257 ± 1.087 |
| Non-food-related | 1.133 ± 1.249 | 1.283 ± 1.012 |
| Placebo test | 0.234 ± 3.303 | -0.031 +- 2.545 |

The very fact that the commercial activity grows faster around Citi Bike stations by itself does not prove that they are the cause, as those stations were not deployed at random, but in the particular locations, important for urban function, which might have already revealed their economic potential. However, looking at the estimated change rates exactly one year prior to deployment, as reported in Tables 2, 3 (placebo test), one can see little change - the average is 0.2%. Looking at the non-food related businesses in similar areas around the deployment (not all of the same census blocks are available in that category, so we included all the available ones), we find much lower growth of 1.1% overall. Therefore, it seems that eating places see a particular growth in terms of both - transaction count as well as sales volume - which happens in exactly the same space and time frame as the Citi Bike deployment. Although it does not provide an ultimate proof of causality within a complex urban system in view of limited insights availability, all the evidence above makes it already highly likely.

For the purpose of additional validation of the observed impact, it would be useful to track how this impact gradually decays with distance from the nearest deployment. Because the granularity and availability of insights does not allow doing this at a sufficient level of detail – we make an attempt by categorizing the blocks by their degree of proximity to the deployment (1-Deployment within the block, 2-Deployment nearby, 3-Deployment further away). However, in Jersey City we are unable to find enough adjacent blocks on different distances from an isolated deployment for such analysis (finding deployments isolated enough for having nearby blocks not affected by other deployments is another challenge).

In order to perform the above analysis, we identify several census blocks affected by the most recent deployment of 2016 in Brooklyn. Because granularity of information does not allow distinguishing eating places, we considered all businesses in the areas of interest. Table 4 shows that the overall change in commercial activity in the identified areas gradually decreases while moving further away from the deployment location. As expected, the overall estimated growth rates are smaller than the growth rates identified for eating places due to the mixed nature of the businesses considered. Statistical significance of the pattern cannot be validated - the difference is not large enough compared to the fluctuations within the control sample.

**Table 4.** Change in commercial activity (all businesses) around most recent deployments in Brooklyn (Sept, 2016) depending on the deployment proximity.

| Buffer zone | Estimated relative change in sales, % | Estimated relative change in transactional activity, % |
| --- | --- | --- |
| 1-Deployment within the block | 1.975 ± 1.310 | 0.987 ± 1.744 |
| 2-Deployment in the neighbor block | 0.700 ± 0.041 | 0.809 ± 0.367 |
| 3-Deployment further away | -0.333 ± 0.518 | 0.122 ± 0.734 |

While the pattern can be seen in commercial activity in the eating places within the close proximity of the Citi Bike deployments based on MRLI data, in order to get a more precise and statistically reliable estimate of the magnitude of such an impact, further analysis based on the individual business earnings/tax revenues or opening/closure rates would be quite helpful if and when such information becomes available from other sources.

**Figure 1.** Commercial activity near Grove Street in August, 2015 (one month before the Citi Bike deployment) against August, 2016 (after the deployment). Screenshots taken from [20]

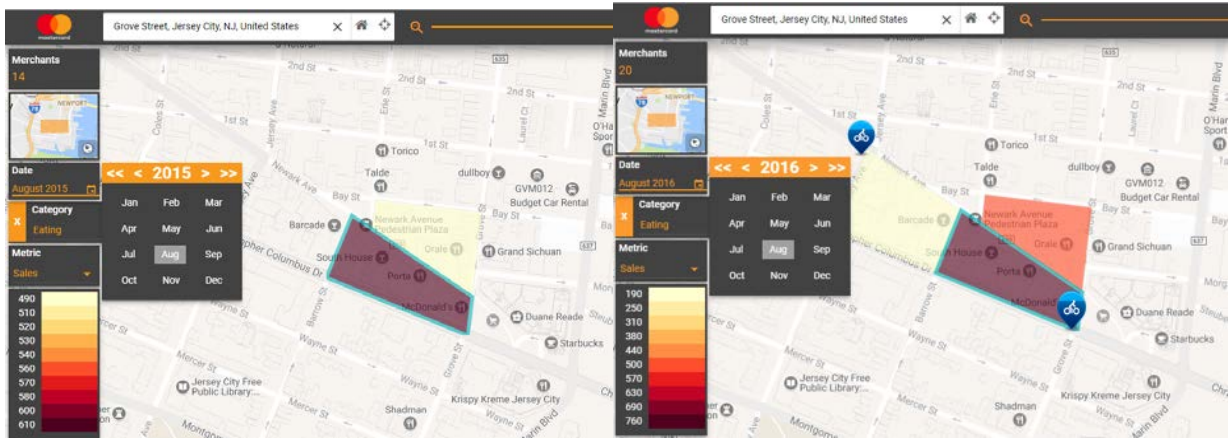

**Impact of the LinkNYC kiosks on the nearby commercial activity**

We expect the LinkNYC deployments (see the visualization of all the deployments on Figure A2 of the Appendix) to intensify the nearby commercial activity by providing visitors with more information about the neighborhood and the available points of sale. Considering the total of 10 LinkNYC deployment areas in Brooklyn (we exclude Manhattan, as many of its locations are among the top percentile by food spending, which makes it more difficult to track fluctuations), we find that only 7 of them have nearby census blocks with reported MRLI metrics, while some have multiple adjacent blocks. The table below reports the full list of the 10 Census Block GeoID's adjacent zip to the Brooklyn LinkNYC locations together with the associated relative changes in transactional activity.

**Table 5.** Census Block GeoID's with available derived information adjacent to the Brooklyn LinkNYC locations and the estimated commercial activity (sales volume) growth rate over the 6 months after the deployment (December 2016 - May 2017) versus similar period one year before (December 2015 - May 2016) and also the growth rate one year before the deployment (December 2015 - May 2016 versus December 2014 - May 2015) for the placebo test.

| Census block GeoID | Estimated relative change after the deployment, % | Estimated relative change one year before, % |
|---|---|---|
| 360470033001001 | 2.8434 | -0.0874 |
| 360470037001000 | -0.9777 | 0.9307 |
| 360470037001009 | 2.8233 | 0.9602 |
| 360470037001010 | 11.4379 | 0.9148 |
| 360470159001000 | 3.18 | 0.302 |
| 360470159005000 | 0.0381 | 1.6165 |
| 360470161003004 | 1.2764 | 0.4064 |
| 360470163003001 | 1.6526 | 0.4071 |
| 360470299001000 | 6.0249 | 1.348 |
| 360470325001001 | 1.9674 | 0.6945 |

The estimated change in total sales volumes vary from -1% to 11%, as can be seen from the table above, with an average at 3.0%. This alone does not mean that this substantial growth has been indeed caused by LinkNYC deployments. However, looking at the change rates exactly before (when the deployment was missing), as reported in the right column of the same table, one cannot see much change - the average is -0.37%.

Considering the 14 zip code areas including those locations, as well as the nearest surrounding zip code areas (see the visualization of the deployment areas in Figure A5 of the Appendix) as a control sample showing the overall trends in the commercial activity across the considered neighborhood, we can find only an average of 0.7% increase. The summary values together with the standard deviations are reported in the table below. We also perform a statistical significance test by evaluating a p-value of the KS-test comparing the growth rate distribution within the areas from deployment and control samples. The p-value, as reported in the summary table, is only 0.29%, i.e. falls considerably below a standard 5% and even a stronger 1% threshold and allows rejecting the hypothesis that the samples follow the same distribution at 99% confidence.

**Table 6.** Estimated average change in commercial activity (all businesses) around most recent LinkNYC deployments in Brooklyn (December 2016), one year before, within the control area and the KS-test results for comparison between the growth distributions within the deployment and control areas.

| Sample | Estimated relative change in sales volume, % | Estimated relative change in transactional activity, % |
|---|---|---|
| Deployment area | 3.027 ± 3.330 | 3.174 ± 2.545 |
| Deployment area one year before | 0.374 ± 1.176 | -0.081 ± 0.782 |
| Control area | 0.721 ± 0.447 | 0.704 ± 0.517 |
| KS-test p-value (deployment vs control) | 0.29% | 0.29% |

While it is still hard to claim the causal relation for certain, as LinkNYC kiosks are not deployed randomly but in specific locations (likely selected as they play an important role in urban function), the evidence above shows a statistically significant impact on commercial activity that we can assume to be associated with LinkNYC deployment. With respect to additionality (compared versus the control area), this gives a 2.3% of additional growth in sales and a 2.4% of growth in transactional activity.

**Discussion**

A causal relationship for indirect impacts such as an impact on commercial activity is always a challenge to prove in complex urban systems [70, 71]. Even though the differences in commercial activity near Citi Bike or LinkNYC deployments turn out to be statistically significant, one might still question if those were the only transformations happening in the area at this time and if there might be a different cause or several causes for the observed effect. Citi Bike stations and LinkNYC kiosks are not placed randomly. The locations are usually picked with respect to multiple considerations, including importance of the location for urban mobility. For example the Citi Bike deployment area in Jersey City contains the main transit hub for commuters into New York City, which is not the case for the control area. There might be other transformative processes specific to the selected locations which might serve as an additional cause of the observed impact. Claiming causality for such indirect impact is always a risk, therefore we frame the conclusions of the present study as a reasonable evidence rather than as a definitive proof of the impact of the above deployments on the local commercial activity.

The temporal "placebo" test showing that the local commercial activity was not going up before the deployment happened further strengthens the above evidence. More detailed data on the commercial activity in individual business locations would be helpful to enable a more detailed analysis, in particular looking at how the impact decays with distance from the deployment location.

Also, when stating the positive impact of the deployments on the local commercial activity one should acknowledge the boundary of the assessment - while there is evidence of growth in commercial activity in the close proximity of the deployments, there is no evidence that this is a pure economic gain or if the increased activity comes at a price of reduced activity in the businesses further away from the deployment due to the competition within commerce. While it is hard to control the dynamics of the commercial activity all over the city due to the limited data coverage, the broader neighborhoods of the deployments considered as a control sample still show an overall growth and the impact is assessed in excess to it.

All said, this present study provides an initial proof-of-concept evaluation of the relation between the commercial activity and the CitiBike and LinkNYC deployments, leaving space for a more detailed follow up analysis if more spatially and temporally granular data becomes available to enable more robust counterfactual analysis and causality evaluation.

**Conclusions**

In this paper we have been able to provide data-driven evidence towards validating the hypothesis that proximity of Citi Bike stations does incentivize local sales at eating places, while LinkNYC kiosks incentivize commercial activity in general. The observed growth of commercial activity near Citi Bike stations and LinkNYC kiosks has been evaluated against the dynamics in different control areas nearby as well as against the dynamics within the deployment areas but before the deployment happened.

Mastercard Retail Location Insights proved to be a new useful source of measurements on the aggregated local commercial activity that can be used in urban analytics, in particular in quantifying economic impact of urban deployments.

**Acknowledgements**
This research was funded by the Future Cities Catapult under the Performance in Use project. The authors would also like to acknowledge support of this research through the data grant from the Mastercard Center for Inclusive Growth for the use of Mastercard Retail Location Insights. We also thank Dr Adam Rae, Dr Claudia Andrade and others from the Future Cities Catapult as well as Achilles Edwin Alfred Saxby, Richard A. Vecsler, Vishwajeet Yashwantrao Shelar, Adrian Dahlin, Maisha Lopa and other collaborators from NYU for their valuable suggestions and discussion.

41(3):260–271.
[17] Paldino S, Bojic I, Sobolevsky S, Ratti C, and Gonzalez MC (2015). Urban magnetism through the lens of geo-tagged photography. *EPJ Data Science*, 4(1):1–17.
[18] Kurkcu A, Ozbay K & Morgul EF (2016). Evaluating the Usability of Geo-located Twitter as a Tool for Human Activity and Mobility Patterns: A Case Study for New York City. *Proceedings of the 95th TRB Annual Conference*, #16-3901, Washington, D.C.
[19] Qian, C., Kats, P., Malinchik, S., Hoffman, M., Kettler, B., Kontokosta, C. and Sobolevsky, S., 2017, July. Geo-Tagged Social Media Data as a Proxy for Urban Mobility. In International Conference on Applied Human Factors and Ergonomics (pp. 29-40). Springer, Cham.
[20] Mastercard Retail Location Insights. https://rliapp.mastercardadvisors.com
[21] Fishman E, Washington S, Haworth N (2014). Bike share's impact on car use: Evidence from the United States, Great Britain, and Australia. *Transportation Research Part D: Transport and Environment*, Volume 31, 2014 (pp. 13-20), ISSN 1361-9209.
[22] Woodcock J, Tainio M, Cheshire J, O'Brien O, Goodman A (2014). Health effects of the London bicycle sharing system: health impact modelling study. *BMJ 2014;348:g425*
[23] Mueller N, Rojas-Rueda D, Cole-Hunter T, De Nazelle A, Dons E, Gerike R, Götschi T, Int Panis L, Kahlmeier S, Nieuwenhuijsen M (2015). Health impact assessment of active transportation: A systematic review. *Preventive Medicine,* Volume 76, 2015 (pp. 103-114), ISSN 0091-7435.
[24] Wang S, Zhang J, Liu L, Duan Z-Y (2010). Bike-Sharing-A new public transportation mode: State of the practice & prospects. *Emergency Management and Management Sciences (ICEMMS)*, 2010 IEEE International Conference on. 222 - 225. 10.1109/ICEMMS.2010.5563463.
[25] Shaheen S, Guzman S, Zhang H (2012). Bikesharing across the Globe. *City Cycling, Urban and Industrial Environments.* 183-210.
[26] NYC DOT Mobility Report, 2016. http://www.nyc.gov/html/dot/downloads/pdf/mobility-report-2016-print.pdf
[27] El- Geneidy A, Van Lierop D & Wasfi R (2015). Do people value bicycle sharing? A multilevel longitudinal analysis capturing the impact of bicycle sharing on residential sales in Montreal, Canada. *Transport Policy.*
[28] Corinne K (2011). Integrating Bike Share Programs into a Sustainable Transportation System. *A product of the National League of Cities, in conjunction with its Sustainability Partner, The Home Depot Foundation.*
[29] The Health Risks and Benefits of Cycling in Urban Environments Compared with Car Use: Health Impact Assessment Study, British Medical Journal, 2011
[30] Andersen LB, Schnohr P, Schroll M, Hein HO (2000). All-cause mortality associated with physical activity during leisure time, work, sports, and cycling to work. *Arch Intern Med.* Jun 12;160(11):1621-8.
[31] Viechnicki P, Khuperkar A, Fishman TD, Eggers WD (2015). Smart mobility: Reducing congestion and fostering faster, greener, and cheaper transportation options. *Deloitte Smart Mobility Research Report.*
[32] Wang X, Lindsey G, Schoner JE, & Harrison A (2012). Modeling Bike Share Station Activity: Effects of Nearby Businesses and Jobs on Trips to and from Stations. *Journal of Urban Planning and Development* 142 (1).
[33] Brooks BA, Schrubbe A (2017). The Need For a Digitally Inclusive Smart City Governance Framework. *UMKC Law Review* 85 UMKC L. Rev. 943
[34] Gupta R, Rao UP (2015). An Exploration to Privacy issues in Location Based Services. *3rd Security*
16

**Appendix**

Figure A1. Citi Bike stations deployment phases in 2013, 2015 and 2016 (visualization created in CARTO).

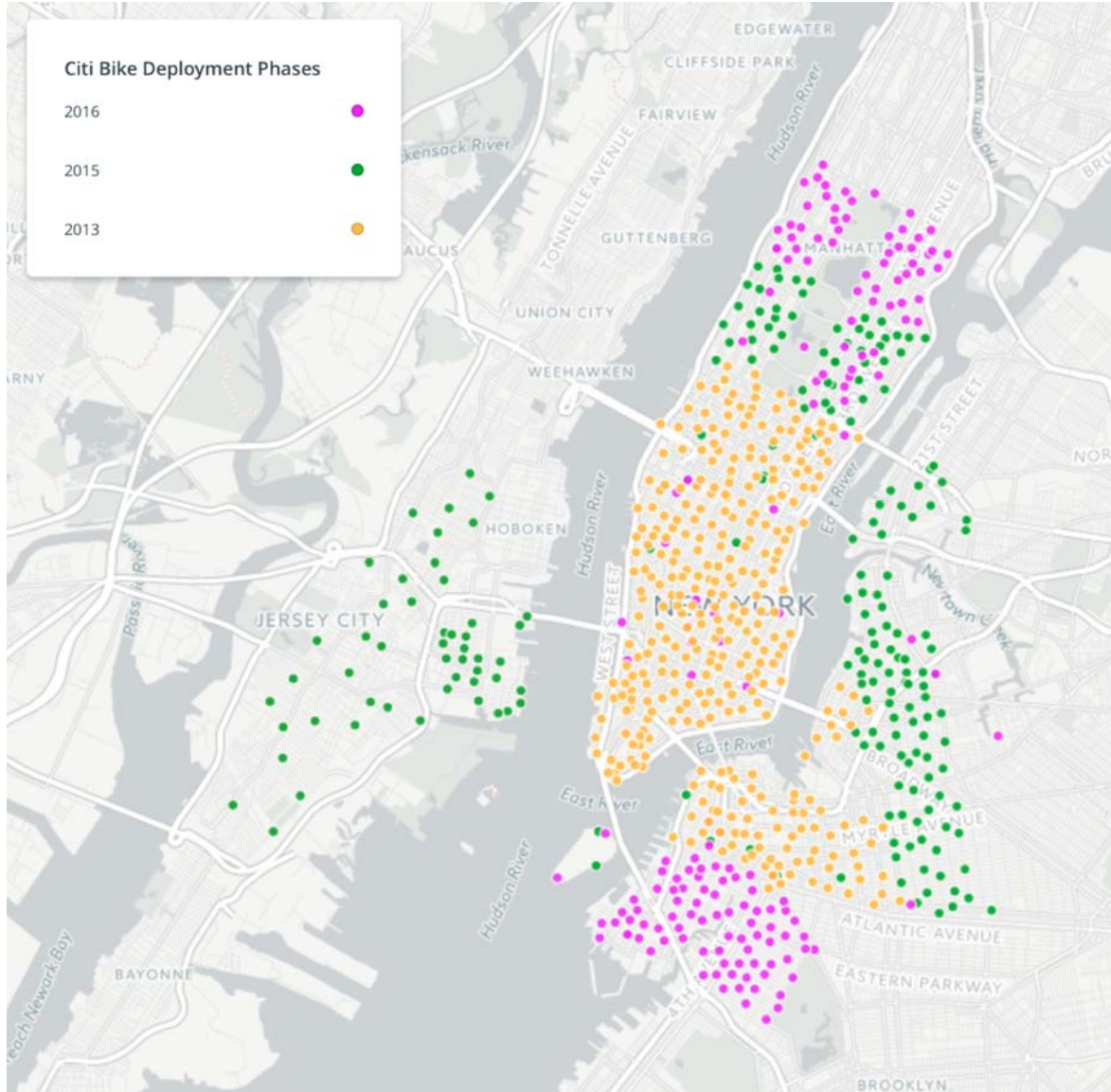

Figure A2. LinkNYC deployments (visualization created in CARTO).

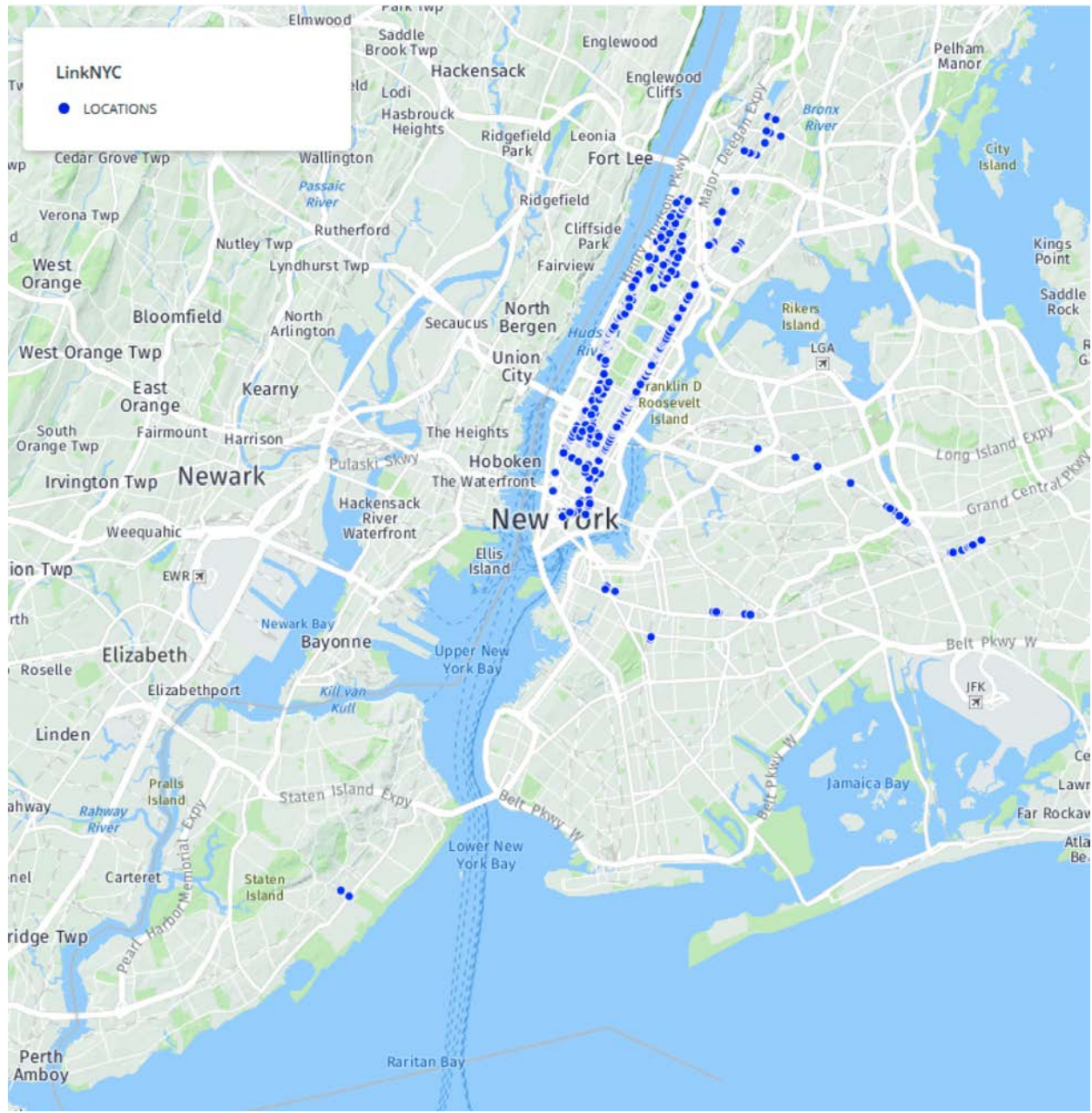

Figure A3. Citi Bike stations deployment phases in 2013, 2015 and 2016 and adjacent zip codes in Brooklyn (visualization created in CARTO).

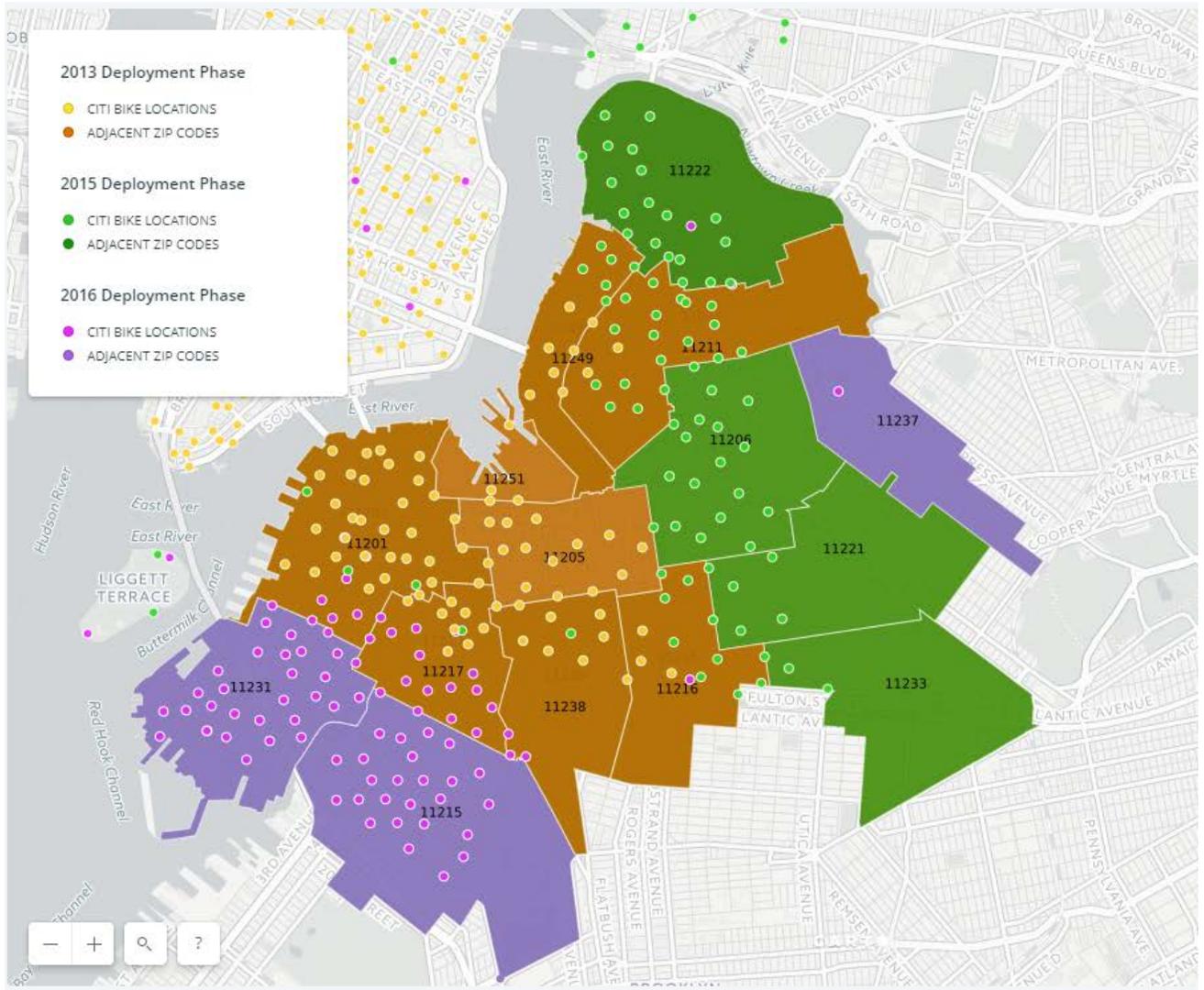

Figure A4. Deployment areas around Journal Square and Grove Street Citi Bike stations in Jersey City - adjacent Census Blocks with data coverage (visualization created in CARTO)

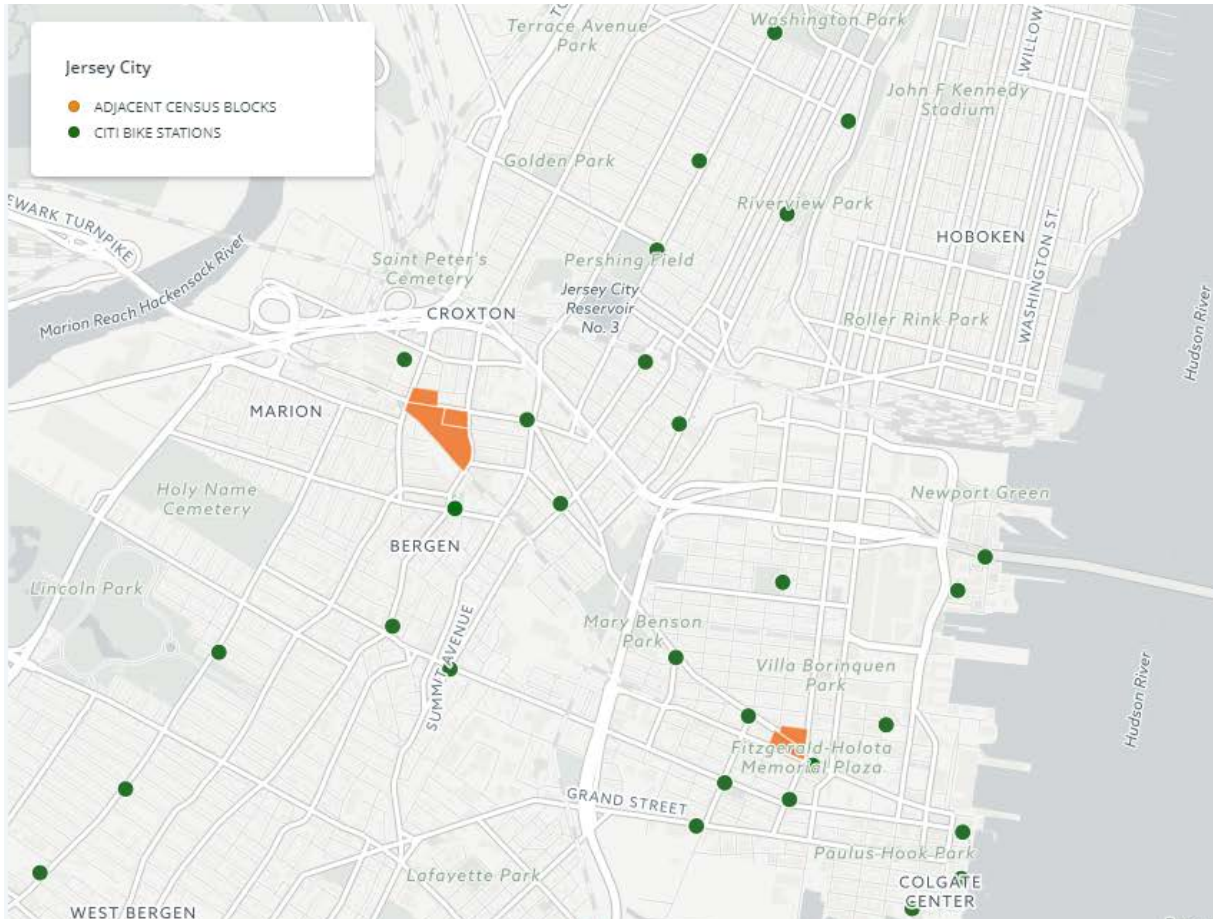

Figure A5. LinkNYC deployment in 2016 in Brooklyn, zip code areas including deployment and control zip code areas nearby (visualization created in CARTO).

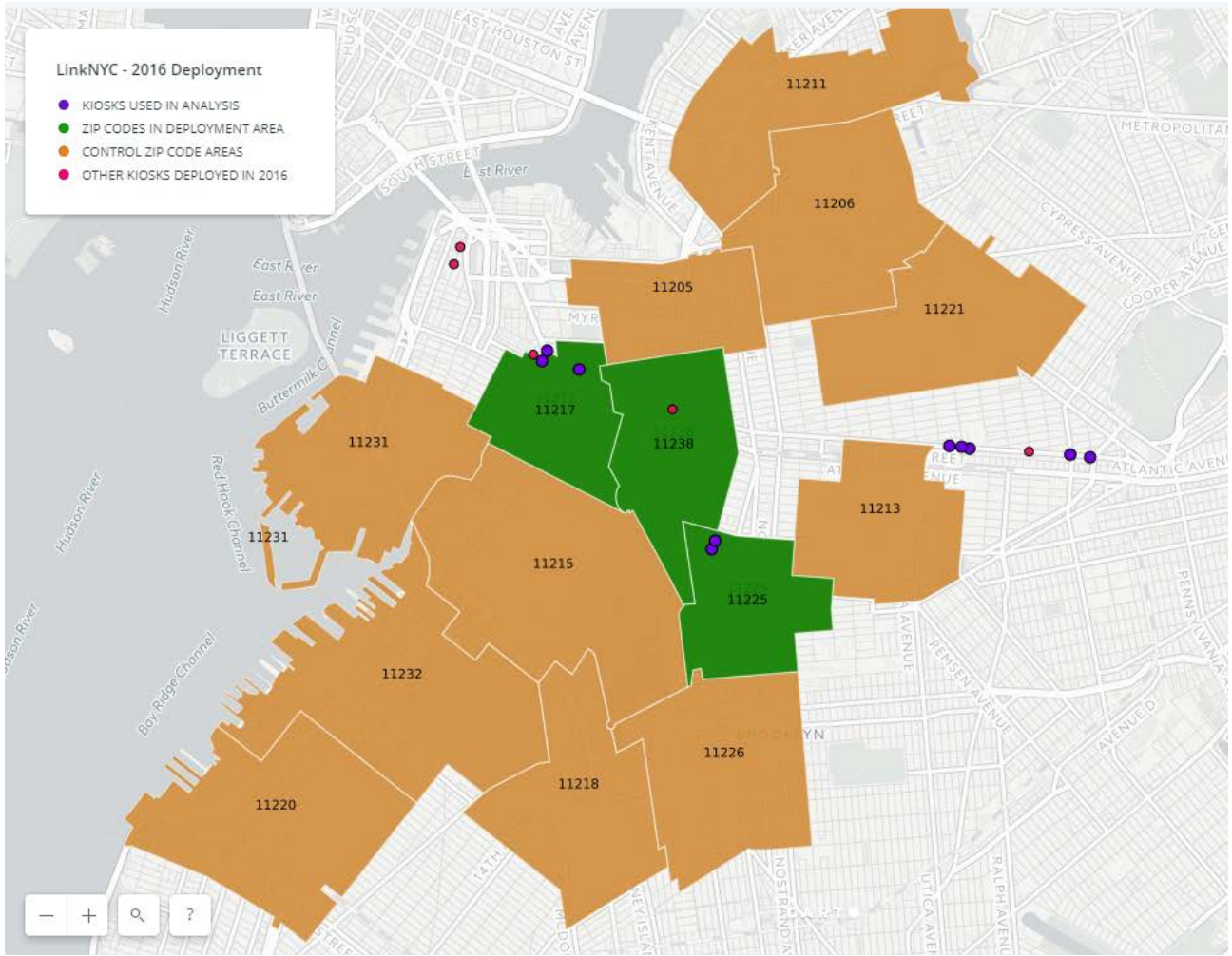